\documentclass[conference]{IEEEtran}

\usepackage{cite}

\usepackage{siunitx} 

\ifCLASSINFOpdf
   \usepackage[pdftex]{graphicx}
\else
\fi
\usepackage{amsmath}
\usepackage{amsfonts} 
\usepackage{array}

\ifCLASSOPTIONcompsoc
 \usepackage[caption=false,font=normalsize,labelfont=sf,textfont=sf,subrefformat=parens,labelformat=parens]{subfig}
\else
 \usepackage[caption=false,font=footnotesize,subrefformat=parens,labelformat=parens]{subfig}
\fi
\usepackage{mathtools}

\usepackage{lineno} 
\def\lc{\left\lceil}   
\def\rc{\right\rceil}
\begin{document}
\title{Towards Differentiable Rendering for Sidescan Sonar Imagery }

%
%
\author{
    \IEEEauthorblockN{
        Yiping Xie,
        Nils Bore,
        John Folkesson
    }
    \IEEEauthorblockA{
        \textit{Robotics, Perception and Learning} \\
        \textit{KTH Royal Institute of Technology} \\
        Stockholm, Sweden \\
        \tt{\footnotesize{\{yipingx, nbore,johnf\}@kth.se}}
    }
}

\maketitle

\begin{abstract}
Recent advances in differentiable rendering, which allow calculating the gradients of 2D pixel values with respect to 3D object models, can be applied to estimation of the model parameters by gradient-based optimization with only 2D supervision. It is easy to incorporate deep neural networks into such an optimization pipeline, allowing the leveraging of deep learning techniques. This also largely reduces the requirement for collecting and annotating 3D data, which is very difficult for applications,
for example when constructing geometry from 2D sensors.
In this work, we propose a differentiable renderer for sidescan sonar imagery. We further demonstrate its ability to solve the inverse problem of directly reconstructing a 3D seafloor mesh from only 2D sidescan sonar data.   
\end{abstract}



%
\IEEEpeerreviewmaketitle

\section{Introduction}
%
%
%


\IEEEPARstart{R}{ecent}  years have seen increasing interest in reconstructing bathymetry from sidescan sonars (SSS) \cite{xie202103}\cite{Bore2021}\cite{xie202112}.
Although multibeam echo sounders (MBES) are the standard in bathymetric surveying from surface ships and large autonomous underwater vehicles (AUVs), the resolution in the across-track direction becomes significantly worse when they are scaled down for smaller AUVs. Sidescan sonars can easily be mounted on small AUVs as they lack the across-track array. However, the disadvantage is that only the ranges of returns are known. The across-track angles are not known. To infer the angles and resolve the ambiguity of 3D reconstruction, one can use data-driven approaches\cite{xie202103}\cite{xie202112} or model-based approaches\cite{Bore2021}. These, however, require accurate navigation estimates, which are generally not available for AUVs due to the cumulative drifts in the dead reckoning.  Besides, data-driven approaches also require 3D supervision to create training data, which generally needs registering sidescan images accurately to available bathymetry (usually from MBES data). All factors mentioned above limit the quality of the reconstructed bathymetry from sidescan. Furthermore, neither highly accurate navigation estimates nor MBES data are available for small AUVs in practice. Nevertheless, if given accurate positioning and accurate measurements for attitudes, recent advances in interferometric sonars with phase differencing techniques can also provide bathymetric measurements provided the sonars mounted on the hull of the survey ships in shallow water.

One solution is to leverage the idea of differentiable rendering\cite{kato2020differentiable} to perform 3D inference with only 2D supervision. Rendering, coming from the field of Computer Graphics, is the process of generating images given camera parameters, geometry, lighting and materials of the 3D scene. In general, rendering algorithms can be divided into rasterization-based approaches and ray tracing-based approaches. Rasterization-based methods are much faster than ray tracing-based methods but provide lower visual quality. Both methods involve non-differentiable operations, such as discrete sampling and path integrals. Differentiable rendering is the technique to tackle the non-differentiabilities in the process of traditional rendering such that the obtained gradients can be used in an end-to-end optimization, making it easy to integrate neural networks into the pipeline. It calculates the gradients of 3D objects and camera properties and then propagates to 2D images, thus allowing optimization of 3D scenes by comparing the rendered images to the measured camera images. Therefore it eliminates the need to collect and annotate 3D data, which is of great benefit for applications. 

\begin{figure}
\centering
\includegraphics[width=3in]{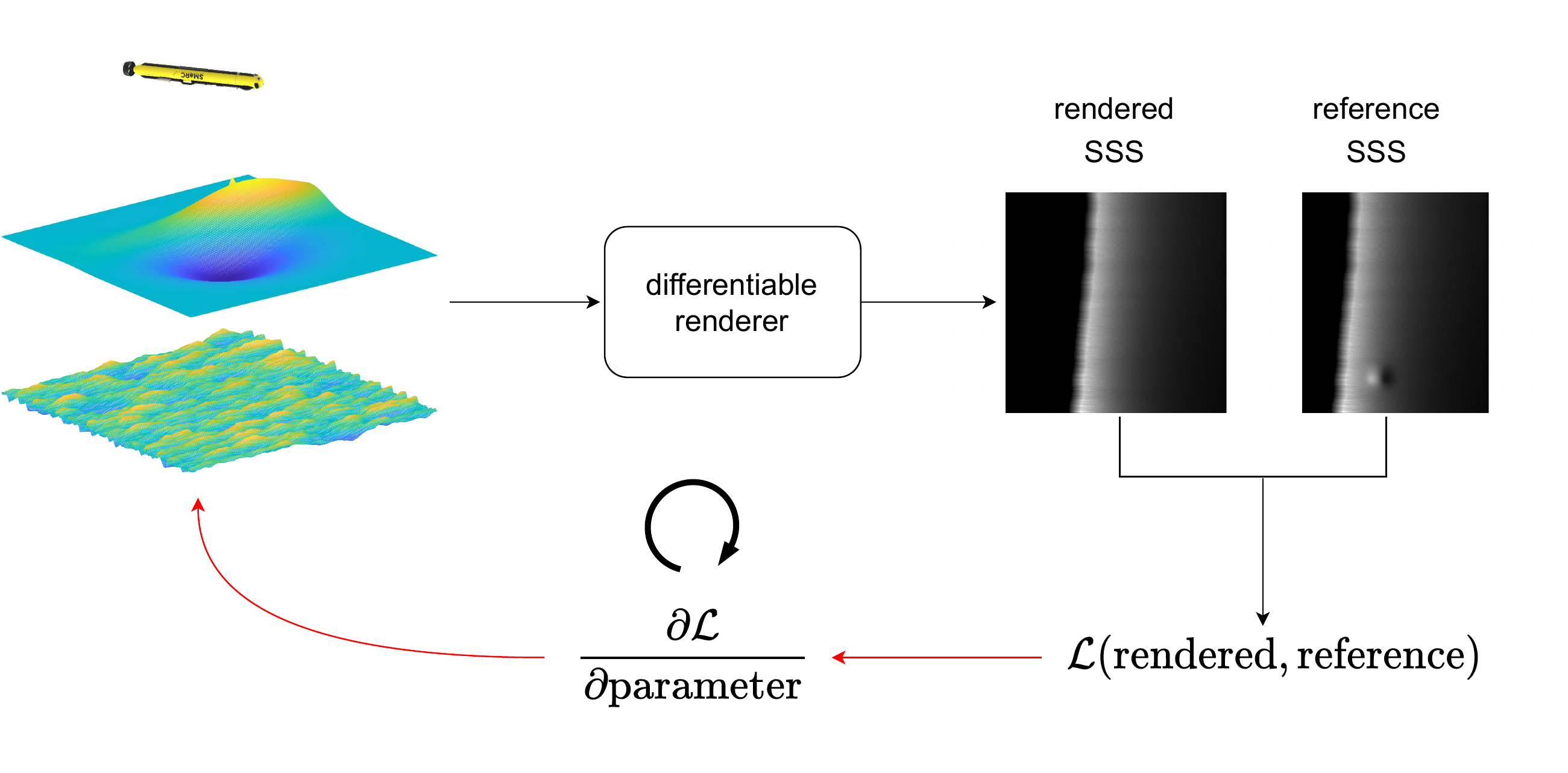}
\caption{Standard schematic overview for a basic optimization using differentiable rendering. The renderer takes sonar's poses, bathymetry mesh and reflectivity to render a SSS image. The supervision is only needed in the form of SSS imagery and the to-be-estimated parameters are updated by backpropagating the error between the reference image and the rendered image. }
\label{fig:gradient_descent_loop}
\end{figure}

There are, in general, two approaches to address the non-differentiabilities in the traditional rendering process. One is to approximate the backward pass of the rendering process, that is, approximate the gradients with respect to input parameters including camera properties, geometry, lighting and textures\cite{kato2018neural}\cite{genova2018unsupervised}\cite{loper2014opendr}\cite{kato2019learning}. The other one is to approximate the forward pass of the rendering such that useful gradients can be obtained\cite{rhodin2015versatile}\cite{liu2019soft}\cite{chen2019learning}. PyTorch3D\cite{ravi2020accelerating} is an open-source library that provides modular differentiable renderers by approximating the rasterization process inspired by  SoftRas\cite{liu2019soft}. Traditional rasterization-based rendering has $xy$- and $z$-discontinuities due to the discrete sampling in rasterization in the $xy$-plane and depth-buffering along the $z$-axis. SoftRas\cite{liu2019soft} proposes to soften the non-differentiabilities by blending contributions of multiple faces for each pixel in a probabilistic manner, both in rasterization and the depth-buffering. PyTorch3D\cite{ravi2020accelerating} decomposes the rendering process into four steps, namely rasterization, lighting, shading, and blending, to make the renderer modular so that we can define our own components according to our needs. In this work, we propose a novel way to define our custom Lambertian shading and Gaussian blending to adapt PyTorch3D's differentiable renderer for cameras to a SSS differentiable renderer. Such a tool for SSS allows  estimation of bathymetry and AUV's poses from supervision only on sonar images by solving a gradient-based optimization problem (see Fig. \ref{fig:gradient_descent_loop}). 

Differentiable rendering can also be used in a deep learning pipeline where some components (e.g., the 3D scene) are replaced by neural networks. There is a rising interest in  using implicit neural representation (INR) and applying neural rendering\cite{tewari2021advances} on a lot of current work on 3D reconstruction. Using implicit neural representations instead of traditional explicit representations for 3D scene  can be beneficial in various ways. For example, the memory requirement with INR is independent of the spatial resolution due to the parameterization of a continuous signal with a neural network, which can be desirable when modeling 3D scene in large scale. 

Our contributions are that:
\begin{enumerate}[\IEEEsetlabelwidth{3)}]
\item We propose a differentiable rendering pipeline for SSS imagery, composed of a soft rasterizer and a SSS shader. The SSS shader applies Lambertian shading and Gaussian blending tailored to SSS's physical process to generate nadir and shadows in SSS images.
\item   We, in this work, demonstrate proof-of-principle use for reconstructing bathymetry from only SSS under the assumption of accurate navigation estimates.
\end{enumerate}
\begin{figure}
\centering
\includegraphics[width=2.2in,angle=270]{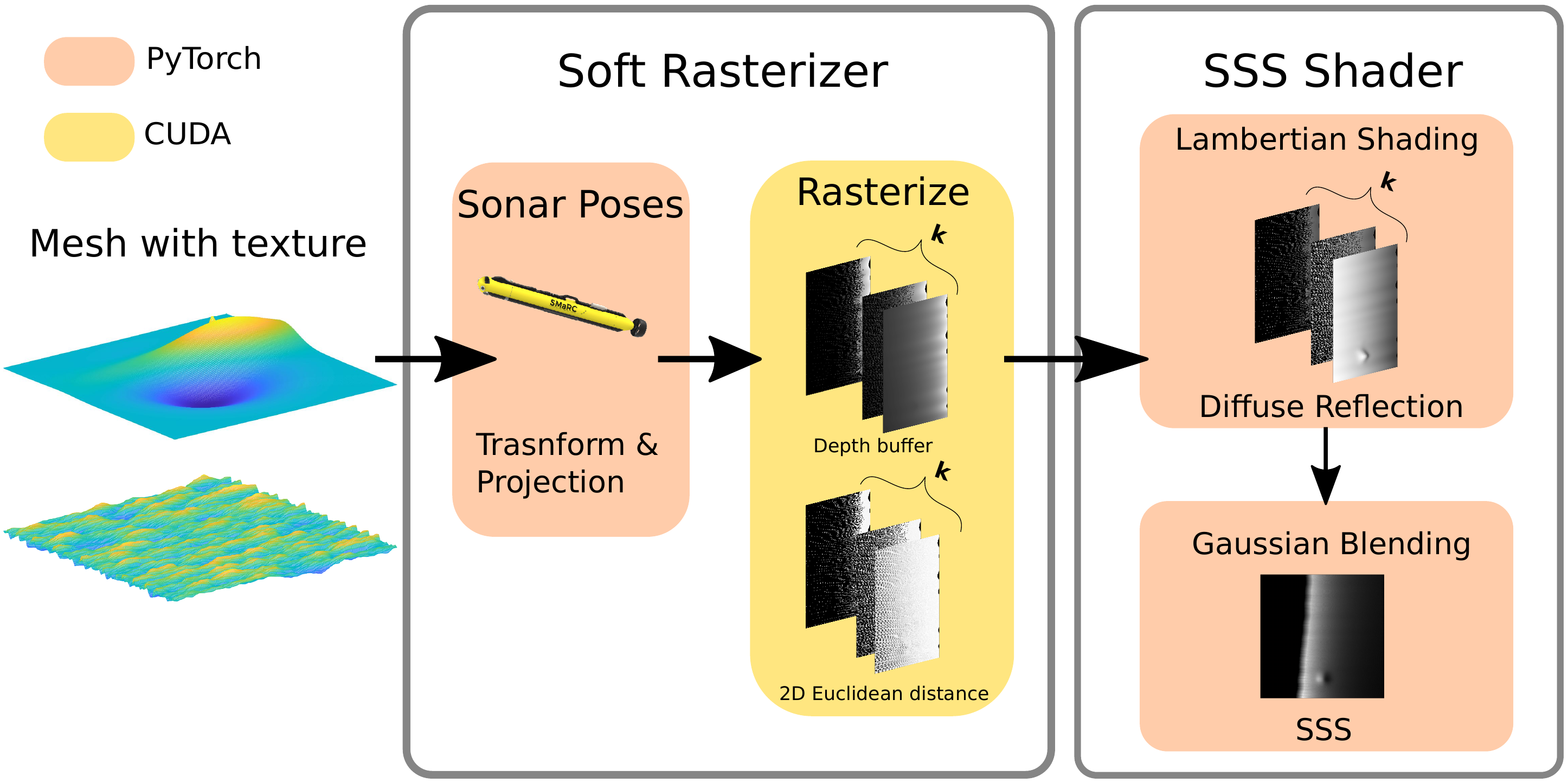}
\caption{The modular SSS imagery rendering pipeline. Given the bathymetry mesh and the reflectivity, the rasterizer returns \textit{Fragment} data\cite{ravi2020accelerating} about the $K$ nearest faces to each pixel based on the sonar's current pose, keeping track of the pixel-to-face distance along the $z$-axis (depth-buffer) and in the $xy$-plane (2D Euclidean distance). Such information is then passed to a SSS shader to apply the lighting, shading and blending.  The lighting and shading are based on the Lambertian model, where only the diffuse reflection is considered. Finally, the Gaussian blending is applied where for each range, a Gaussian kernel centered at that range is used to weight the contribution across ranges from the sonar. }
\label{fig:component}
\end{figure}
\subsection{Related Work}

\subsubsection{Non-differentiable SSS simulators}
Several non-differentiable SSS simulators based on ray tracing methods\cite{bell1997simulation}\cite{riordan2005implementation}\cite{pailhas2009real} and pseudospectral time-domain (PSTD) methods\cite{elston2004pseudospectral} 
have been developed over the last few decades. These methods include accurately simulating the physical process, reproducing important features of SSS, e.g., random speckle and shadows, generating realistic SSS images and even simulating in real-time\cite{riordan2005implementation}\cite{pailhas2009real}. The applications of such simulators are testing the effectiveness of image mosaicing, detection and classification algorithms, modeling lawnmower pattern surveys, investigating vehicle position estimation and etc.

\subsubsection{Differentiable rendering}
There are several open-source libraries that support differentiable rendering for camera images and CUDA implementations. As for the rendering methods, PyTorch3D\cite{ravi2020accelerating}, Kaolin\cite{jatavallabhula2019kaolin} and TensorFlow Graphics\cite{valentin2019tensorflow} are based on rasterization while Mitsuba 2\cite{nimier2019mitsuba} is based on ray tracing. Ray tracing-based methods can generate more photo-realistic rendering but come with the cost of requiring more computation resources. Thus, for applications where visual quality is not the primary objective, rasterization-based methods are more suitable. Among rasterization-based libraries, PyTorch3D provides a modular rendering pipeline composed of a rasterizer module and a shader module (Fig. \ref{fig:component}). In the rendering pipeline of PyTorch3D, the rasterizer first applies 3D coordinate transformation and then rasterization based on the provided primitives (e.g., triangle meshes). The output is afterwards passed to the shader module to apply lighting, texturing, shading and blending to generate the final image. Such a customization-oriented design makes it easy to extend the sub-modules. For example, \cite{wilmanski2022differentiable} replaces the shader in PyTorch3D with a neural shader implemented as a conditional generative adversarial network (cGAN) to generate realistic synthetic aperture radar (SAR) images in a differentiable manner. Similarly, one can in theory use the cGAN in \cite{Bore2020} to model the scattering effect for SSS as well. 
Another example is to use implicit neural representation where a neural network represents the 3D scene and/or objects, such as Occupancy Networks \cite{mescheder2019occupancy}, Deep Signed Distance Function (DeepSDF) \cite{park2019deepsdf}, Scene Representation Networks (SRNs) \cite{srn2019}, Neural Radiance Fields (NeRF) \cite{nerf2020}, and Sinusoidal Representation Networks(SIRENs)\cite{siren2020}. \cite{Bore2021} and \cite{xie202112} both use SIRENs to represent the bathymetric map and optimize the self-consistent bathymetry from sidescan and sparse altimeter readings.

\section{Method}
\subsection{Sidescan Sonar Formation}
\begin{figure}[t]
\centering
\includegraphics[width=3.1in]{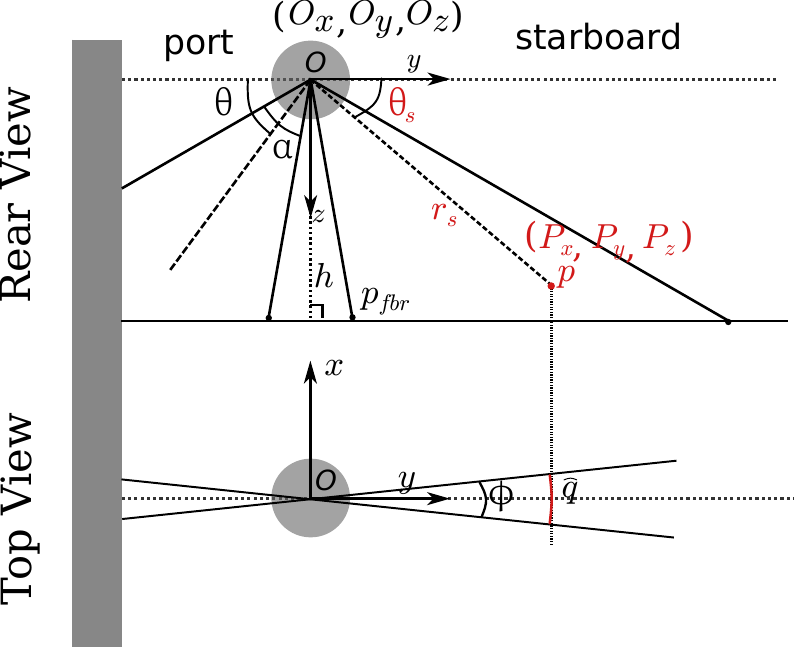}
\caption{Sidescan sonar formation in Forward, Lateral Down (FLD) frame. The sidescan has two heads, port and starboard, symmetrically placed to each side with a fixed \textit{tilt angle}. Usually, the sonar has a wide vertical beam angle $\alpha$ and a very narrow horizontal beam angle $\phi$. The wide $\alpha$ allows sidescan to have a wide swath and the narrow $\phi$ gives sidescan imagery high resolution in the across-track direction. The first meaningful return reflected from the seabed $p_{fbr}$ is called \textit{first bottom return}, the detection of which is very useful to determine the sonar's altitude $h$.}
\label{fig:sss-formation}
\end{figure}

Fig. \ref{fig:sss-formation} illustrates the top view and the rear view of a sidescan sonar with its sensor origin $O$ at altitude $h$, with coordinates $(O_x,O_y,O_z)$ in the world coordinate frame. The vertical beam width $\alpha$, sometimes referred to as sensor opening in the yz plane, is usually $40\text{-}60^\circ$. The sonar is mounted with a \textit{tilt angle} $\theta$, which is the angle between the horizontal axis and the center of the vertical beam width $\alpha$. The horizontal beam width $\phi$, sometimes referred to as sensor opening in the $xy$-plane, is usually around $1^\circ$. The beam pattern of $\phi$ can be modeled based on the theoretical beam pattern of a linear phased array. 

Let $p$ (with coordinates $(P_x,P_y,P_z)$ in the world coordinate frame) be the footprint in the ensonified region on the bathymetric surface, whose polar coordinates can be expressed in its slant range $r_s$ and its grazing angle $\theta_s$. The grazing angle $\theta_s$ can be calculated as follows if we ignore ray bending effects and $P_z$ is known:
\begin{equation}\label{eq:grazing_angle}
    \theta_s = \arcsin{\Big(\frac{|P_z-O_z|}{r_s}\Big)}.
\end{equation}
However, in reality, the bathymetry is usually unknown, thus leaving ambiguity in $\theta_s$. Due to the horizontal beam width $\phi$, the exact point position of $p$ in the $xy$ plane is also ambiguous over the arc $q$ as shown in Fig. \ref{fig:sss-formation}.
\subsection{Differentiable Rendering}
\subsubsection{Sonar Model vs Camera Model}
We model the sidescan sonar as an acoustic camera where for each ping, the sonar's beam center is the camera's optical center. The extrinsic parameters are calculated based on the sonar's current pose.
Here we use a perspective camera model with a horizontal field of view (FoV) $\phi$ and a vertical FoV $\alpha$ by adjusting the image aspect ratio, where the intrinsic parameters are:
\begin{equation}
    K_{in}=\begin{bmatrix}
f_x & 0 & 0 & 0\\
0 & f_y & 0 & 0\\
0 & 0 & 0 & 1\\
0 & 0 & 1 & 0
\end{bmatrix},f_x=\frac{1}{tan(\frac{\phi}{2})},f_y=\frac{1}{tan(\frac{\alpha}{2})}.
\end{equation}
The horizontal beam pattern is modeled as:
\begin{equation}
    bp_{\phi}=\bigg(\frac{k_{\phi}\cdot sin(\phi-\phi_0)}{sin(k_{\phi}\cdot sin(\phi-\phi_0))}\bigg)^4,
\end{equation}
with $\phi_0=0.5^\circ$ and $k_{\phi}=159.46$ for a (one-way) 3 dB beam width of $\phi=1^\circ$, similar as in \cite{folkesson20}.
\subsubsection{Soft Rasterizer}
The size of the rendered camera image can be set accordingly, e.g., as $\lc \frac{f_x}{f_y}\cdot W \rc \times W$. The rendered \textit{Fragment} data for one ping of sidescan will contain the rendered features from the nearest $K$ mesh faces for each pixel on the camera image plane, which is determined by the closest top $K$ faces that are intersected with the current ray based on the depth-buffer. These features (see Fig. \ref{fig:component}) are going to be blended where the influence of multiple faces depends on the pixel-to-face distance along the $z$-axis and in the $xy$-plane.

\subsubsection{SSS Shader}
The lighting and shading are modeled based on the Lambertian cosine law, where only the diffuse reflection is computed, where the light is considered as a point light at the sonar's location. The blending is to combine influence from $K$ faces into one final value. When blending in the $xy$-plane, the same as \cite{ravi2020accelerating}, we apply a sigmoid function controlled by a hyperparameter $\sigma$ to blur the face boundaries based on the distance in the $xy$-plane.
In each ping of SSS, the returned intensity for each bin comes from a specific range interval, i.e., the slant range $[r_s-\frac{\Delta r}{2},r_s+\frac{\Delta r}{2}]$, under the assumption of isovelocity. Therefore, when blending along the $z$-axis for each bin, we apply a Gaussian kernel centered at the corresponding slant range $r_s$:
\begin{equation}
    k_g = e^{\frac{-(r-r_s)^2}{\gamma}},
\end{equation}
where $r$ is obtained from the depth-buffer and $\gamma$ is a hyperparameter to control the scaling of the exponential function defined by the user. After applying the Gaussian kernel $k_g$, all returns along the arc (at the corresponding range) are summed to compute the sidescan intensity for each bin to simulate the sidescan's physical process. The SSS intensity $I_{r_s}$ for range $r_s$ is blended by summing all the returned intensities within the FoV over top $K$ nearest faces: \begin{equation}\label{eq:naive_blending}
    I_{r_s}= \sum_{1}^K\sum_{\phi_{\textrm{min}}}^{\phi_{\textrm{max}}}\sum_{\alpha_{\textrm{min}}}^{\alpha_{\textrm{max}}} {k_g \cdot \textrm{sigmoid}\big(\frac{d_{xy}}{\sigma}\big) \cdot c},
\end{equation}
where $d_{xy}$ is the 2D pixel-to-face distances in the $xy$-plane and $c$ is the intensities for each of the top $K$ faces per pixel which is the output of the Lambertian shader. 

Such a Gaussian blending models the physical process of sidescan and softens the non-differentiabilities of such process at the same time. Note that not only the shadows are naturally rendered but also the nadir by the Gaussian blending.
\subsubsection{Layover Phenomenon}
\begin{figure*}[h!]
\centering
\includegraphics[width=7.1in]{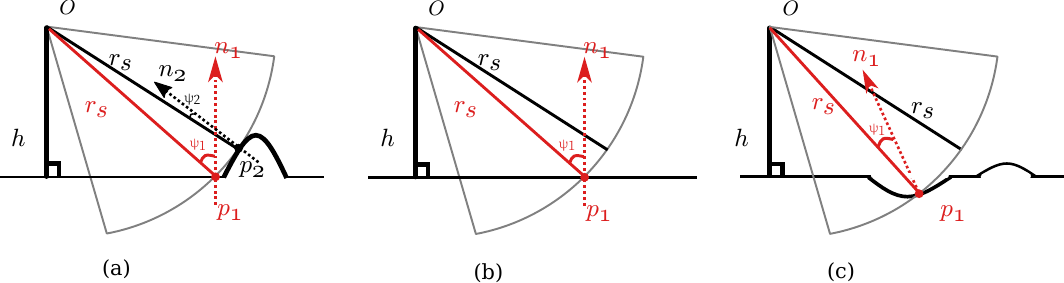}
\caption{An example of the layover phenomenon. (a) Echos from two parts of the seafloor $p_1$ and $p_2$ arrive simultaneously at the sensor $O$. (b) Before optimizing to reconstruct the shape of the seafloor, we normally use flat seafloor assumption for initialization. (c) At the early stages of the optimization, there will be a hole around $p_1$ compensated for the layover phenomenon.}
\label{fig:ambiguity}
\end{figure*}
Note that for each bin of sidescan, since we only know the slant range but not the grazing angle, we are essentially summing returns from all the angles (within the vertical beam width) within that range interval, where most of the returns are zeros due to they are corresponding to returns from water columns. However, in certain circumstances, there could be multiple surfaces where they are at the same distance from the sonar and the returns would be added up to a higher intensity, which is sometimes referred to as \textit{layover} \cite{woock2010deep}. In this case, they will be multiple non-zero returns from multiple angles for one sidescan bin. 

Such layover phenomenon and angles ambiguity would cause problems at the early stage of the optimization, where we try to reconstruct the bathymetry from sidescan images with the differentiable rendering pipeline (Fig. \ref{fig:gradient_descent_loop}). An example is shown in Fig. \ref{fig:ambiguity}, where as a ground truth [Fig. \ref{fig:ambiguity} (a)] there are two points on the seafloor $p_1$ and $p_2$ that are at the distance $r_s$ from sonar at $O$. Usually, we need to initialize the bathymetry with a flat seafloor [Fig. \ref{fig:ambiguity} (b)] according to the sonar's altitudes before the optimization starts. The ground truth sidescan intensity for that bin at that range $r_s$ actually comes from the sum of reflection from $p_1$ with incidence angle $\psi_1$ and $p_2$ with incidence angle $\psi_2$, which is unfortunately not known to the optimizer. This will result in the optimizer trying to move the vertices both around $p_1$ and $p_2$ to compensate for the higher intensity value [Fig. \ref{fig:ambiguity} (c)]. Observations from other views at the same region and optimizing long enough until convergence could help, but we propose summing the top $M$ largest returns along the arc instead of summing them all as in \eqref{eq:naive_blending}. The reason that it would help speed up the convergence in this situation is that $\psi_2<\psi_1$, meaning $|cos\psi_2|>|cos\psi_1|$, and ignoring smaller-value returns when computing the intensity would reduce the gradients with respect to vertices at $p_1$ being propagated back.

\section{Experiments}
\label{par:experiments}
\begin{figure}[h]
\centering
\subfloat[GT dome mesh]{\includegraphics[width=1.25in]{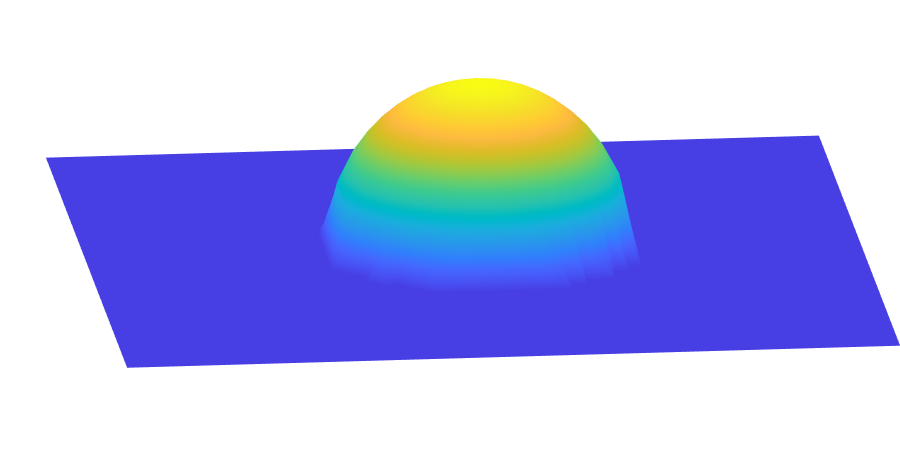}\label{fig:dome_mesh}}\quad
\subfloat[SSS lines ]{\includegraphics[width=0.6in]{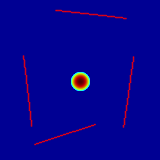}\label{fig:dome_sss}}\quad
\subfloat[Reconstructed]{\includegraphics[width=1.25in]{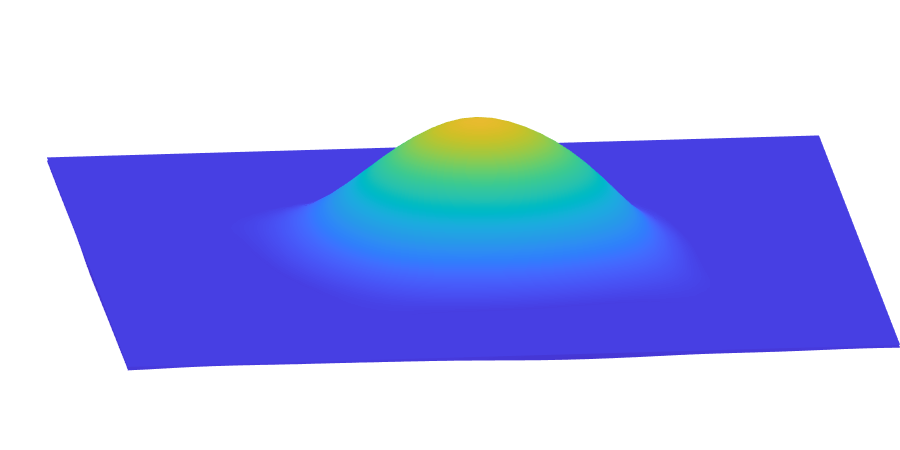}\label{fig:dome_hat_mesh}}
\caption{Results of reconstructing a dome on flat seafloor from reference SSS only. (a) Ground truth (GT) dome mesh with a 5 meter radius. (b) The simulated SSS survey to "see" the dome from different viewpoints. (c) The reconstructed dome after optimization where the top is at 16.12 meter depth.}
\label{fig:toy_example_results}
\end{figure}

\begin{figure}[h]
\centering
\includegraphics[width=3.5in]{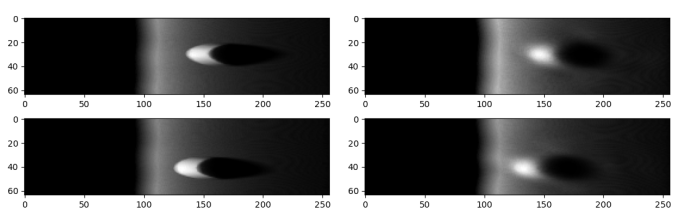}
\caption{Examples of SSS images rendered from GT dome (left) and reconstructed dome (right). We can see from the SSS that the dome, the shadow and the nadir area are very well recovered.}
\label{fig:toy_sss}
\end{figure}
We demonstrate two experiments to verify if the proposed differentiable renderer can be used to reconstruct bathymetry with only SSS images at different view angles. One is a toy example where the ground truth bathymetry is a dome on a flat seafloor. The other experiment is from real data, where the collected MBES data is used to construct the ground truth bathymetry mesh with 0.5 meter resolution. In both experiments, we use the ground truth bathymetry and the differentiable renderer to simulate the reference SSS. Also in both cases, we simulate the SSS with the same settings: horizontal beam width $\alpha=1^\circ$, vertical beam width $\phi=50^\circ$, maximum slant range 50 meters, mounted at a tilt angle $\theta=30^\circ$. Each ping of sidescan has 256 bins, which gives us $\sim0.2$ meter across-track resolution. In addition, we use the plain textures of mesh, assuming the reflectivity of the seabed is constant in both experiments. 
\begin{figure*}[!h]
\centering
\subfloat[GT mesh of the rocky area]{\includegraphics[width=2.2in]{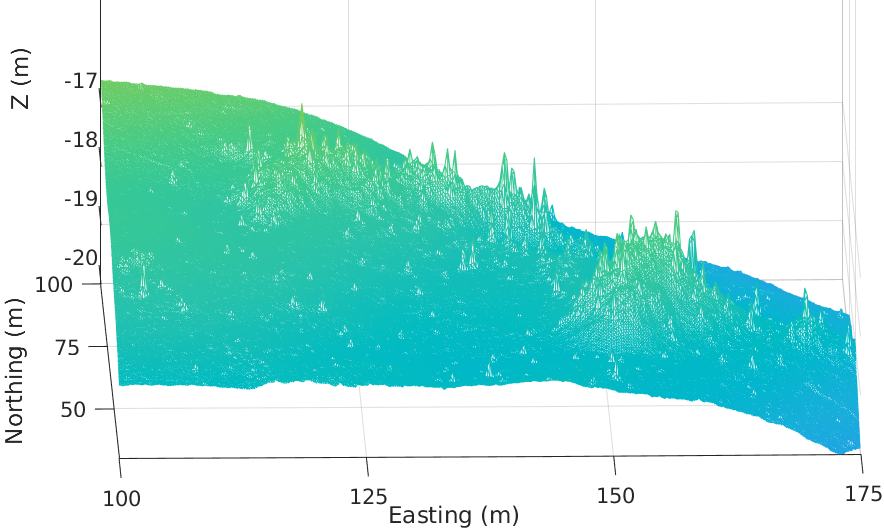}\label{fig:real_mesh}}\quad
\subfloat[Initialized mesh before optimization]{\includegraphics[width=2.2in]{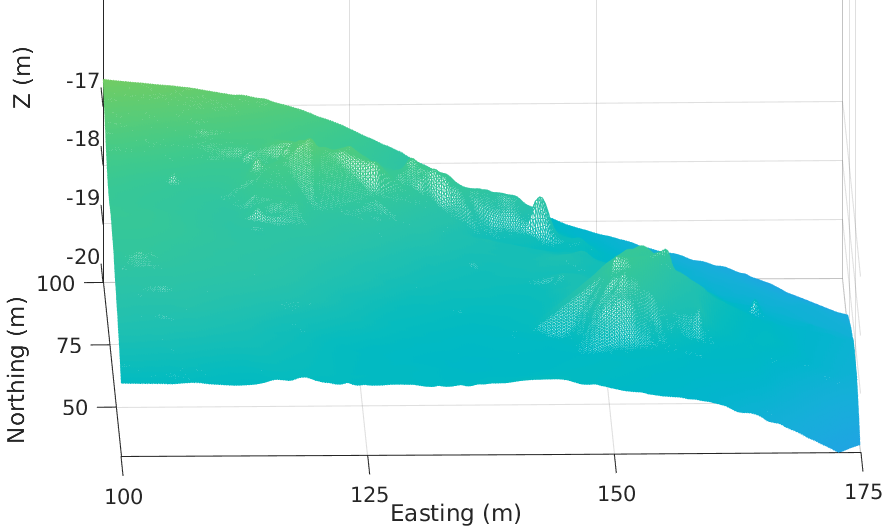}\label{fig:real_mesh_init}}\quad
\subfloat[Reconstructed mesh after optimization]{\includegraphics[width=2.2in]{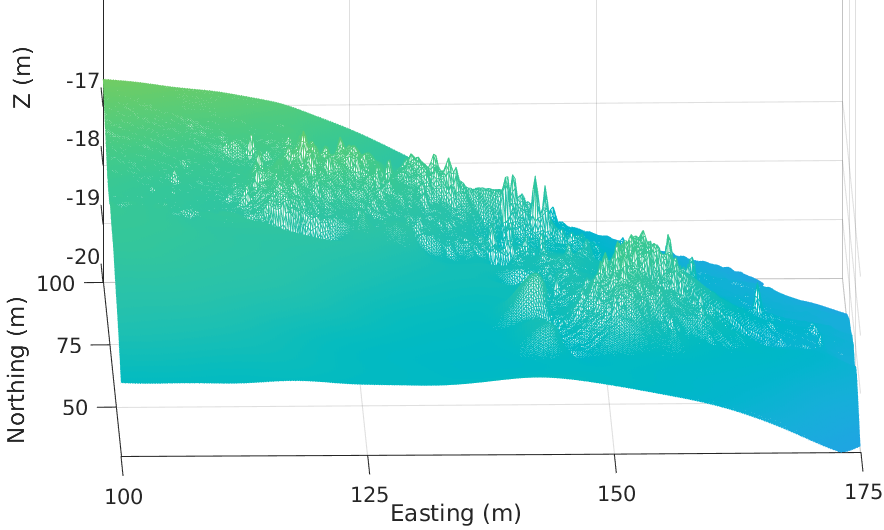}\label{fig:real_mesh_hat}}
\caption{Results of reconstructing rocky area in shallow water from reference SSS only. (a) Ground truth mesh (75m x 50m) constructed from MBES data. (b) Initialized mesh from altimeter readings after interpolation and smoothing. (c) Reconstructed mesh from 3 lines of SSS after 100 epochs of optimization.}
\label{fig:real_mesh_results}
\end{figure*}
\subsubsection{Toy Example}
We simulate the bathymetry as a dome with radius of 5 meters on the flat seafloor at 20 meter depth [Fig. \subref*{fig:dome_mesh}] as ground truth. To generate reference SSS images, we simulate 4 SSS lines around the dome, as shown in  Fig. \subref*{fig:dome_sss}, which are used as the ground truth SSS (Fig. \ref{fig:toy_sss} left) for the bathymetry reconstruction. We start with a complete flat seafloor at 20 meter depth and render the initial SSS images given sonar's poses and initial bathymetry mesh. We optimize the mesh by computing the loss between current SSS images and the reference SSS and a regularization loss on the mesh, which is the normal consistency regularization as implemented in \cite{ravi2020accelerating}. The normal consistency regularization encourages neighboring faces to have similar surface normal directions. 

We train 100 epochs and the estimated bathymetry is as shown in Fig. \subref*{fig:dome_hat_mesh}, where the top of the dome is at 16.12 meter depth. And we can also see the reconstructed dome from the rendered sidescan images [Fig. \subref*{fig:dome_sss}]. Because we use a dome sitting on the flat seafloor as the ground truth, the transition from the flat seafloor to the dome is not smooth at all. We can see from Fig. \subref*{fig:dome_hat_mesh} that the transition is much smoother mainly due to the normal consistency regularization on the mesh.

\subsubsection{Real Dataset}
\begin{figure}[t]
\centering
\subfloat[GT SSS]{\includegraphics[width=1.34in]{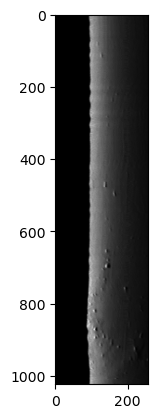}\label{fig:real_sss_gt}}\quad
\subfloat[SSS rendered ]{\includegraphics[width=1.34in]{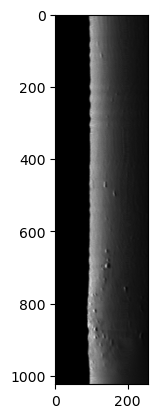}\label{fig:real_sss_hat}}\quad
\caption{An examples of SSS images containing 1024 pings rendered from the ground truth real-world bathymetry (left) and rendered from the estimated bathymetry (right). }

\label{fig:sss_example_results}
\end{figure}

We use the bathymetry formed from the collected MBES data from a survey ship at a shallow area (depth at 9-25 meters) as ground truth. The survey ship  is equipped with an RTK GPS to ensure high positioning. The sonar's poses from 3 lines at a rocky area [see Fig. \subref*{fig:real_mesh}] are used to generate reference SSS images given the GT bathymetry. An example of the reference SSS images is from Fig. \subref*{fig:real_sss_gt}, where we can clearly see many rocks from the SSS.

We construct the sparse depth from altimeter readings and pressure sensor readings and apply linear interpolation between the sparse depth and then Gaussian smoothing to initialize the bathymetry mesh.[see Fig. \subref*{fig:real_mesh_init}]. We can notice that the initialized seafloor has the same large-scale features as the ground truth but no small features such as small rocks. Again we optimize the mesh by computing the loss between rendered SSS images given current bathymetry and the reference SSS and the normal consistency regularization for 100 epochs. The reconstructed seafloor can be seen in Fig. \subref*{fig:real_mesh_hat}, where we can clearly see that many rocks are reconstructed. Qualitatively, the estimated bathymetry has shown high-quality reconstruction with reproducing notable small features. In addition, we can see the rocks from the rendered SSS [Fig. \subref*{fig:real_sss_hat}] as well.

\section{Discussion}\label{par:discussion}
Note that in the previous work of reconstructing bathymetry from SSS\cite{xie202103}\cite{Bore2021}\cite{xie202112}, constraints from sparse depth need to be explicitly modeled during the optimization/training. The sparse depth can be from altimeter readings or first bottom return detection from the sidescan data itself. The former would introduce errors from altimeter sensors, while the latter would introduce errors from the first bottom return detection algorithm. However, using the differentiable rendering pipeline proposed in this work, the information contained at the nadir area is implicitly modeled in the SSS rendering process, which will avoid the errors mentioned above and gives a more elegant method of constraining the bathymetric solution. Besides, the proposed method can utilize the information from the shadows in SSS images to infer the properties of the objects causing the shadows, which the previous work \cite{xie202103}\cite{Bore2021}\cite{xie202112} were not able to do. Such a differentiable renderer for SSS shows the potential to reconstruct bathymetry with small details, e.g., the small rocks in Fig. \subref*{fig:real_mesh_hat}.

\section{Conclusion and future work}

We present a differentiable renderer for sidescan imagery and successfully demonstrate that it can be of great use to reconstruct bathymetry from sidescan images only. However, there are a few parts needed to improve in the future to make this tool more useful for real-world applications. One is to find a way to include the reflectivity estimation in the differentiable rendering pipeline, which in theory is not difficult if we can have a reasonable guess of the reflectivity even in a coarse-scale. Another thing is to use the real SSS collected from the sonar as the reference, which requires either the modeling of speckle noise in the raw SSS images due to the scattering process or denoising the real SSS images before computing the loss. Finally, we would like to estimate the sonar's poses at the same time as reconstructing the bathymetry, which is most important when we want to reconstruct super-resolution bathymetry leveraging the high across-track resolution of the sidescan.


%



\section*{Acknowledgment}
This work was supported by the Wallenberg AI, Autonomous
Systems and Software Program (WASP) and  by the Stiftelsen  för  Strategisk  Forskning (SSF)  through  the  Swedish  Maritime  Robotics  Centre  (SMaRC)(IRC15-0046). Our dataset was acquired in collaboration with MarinMätteknik (MMT) Gothenburg. 

The computations were enabled by resources provided by the Swedish National Infrastructure for Computing (SNIC) at Chalmers Centre for Computational Science and Engineering (C3SE) partially funded by the Swedish Research Council through grant agreement no. 2018-05973.
\ifCLASSOPTIONcaptionsoff
  \newpage
\fi



\bibliographystyle{ieeetr}
\bibliography{IEEEfull,ieeeref}
%



%

\begin{IEEEbiography}[{\includegraphics[width=1in,height=1.25in,clip,keepaspectratio]{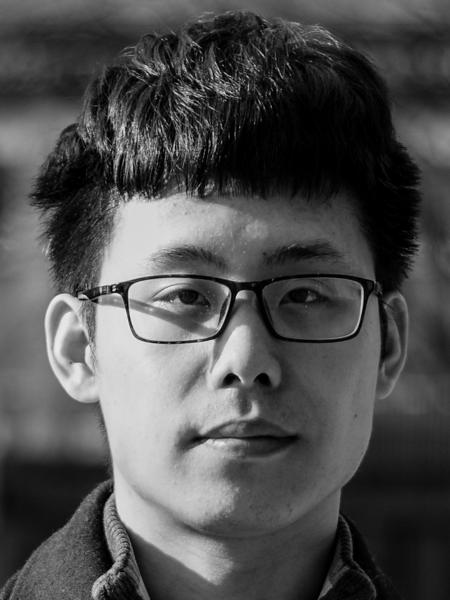}}]{Yiping Xie}
received the B.S. degree in electrical engineering from Beihang University, Beijing, China, in 2017, and the M.Sc. degree in computer science from Royal Institute of Technology (KTH), Stockholm, Sweden, in 2019. 

He is currently a Ph.D. student with the Wallenberg AI, Autonomous Systems and Software Program (WASP) from the Robotics Perception and Learning Lab at KTH. His research interests include perception for underwater robots, bathymetric mapping and localization with sidescan sonar.
\end{IEEEbiography}

\begin{IEEEbiography}[{\includegraphics[width=1in,height=1.25in,clip,keepaspectratio]{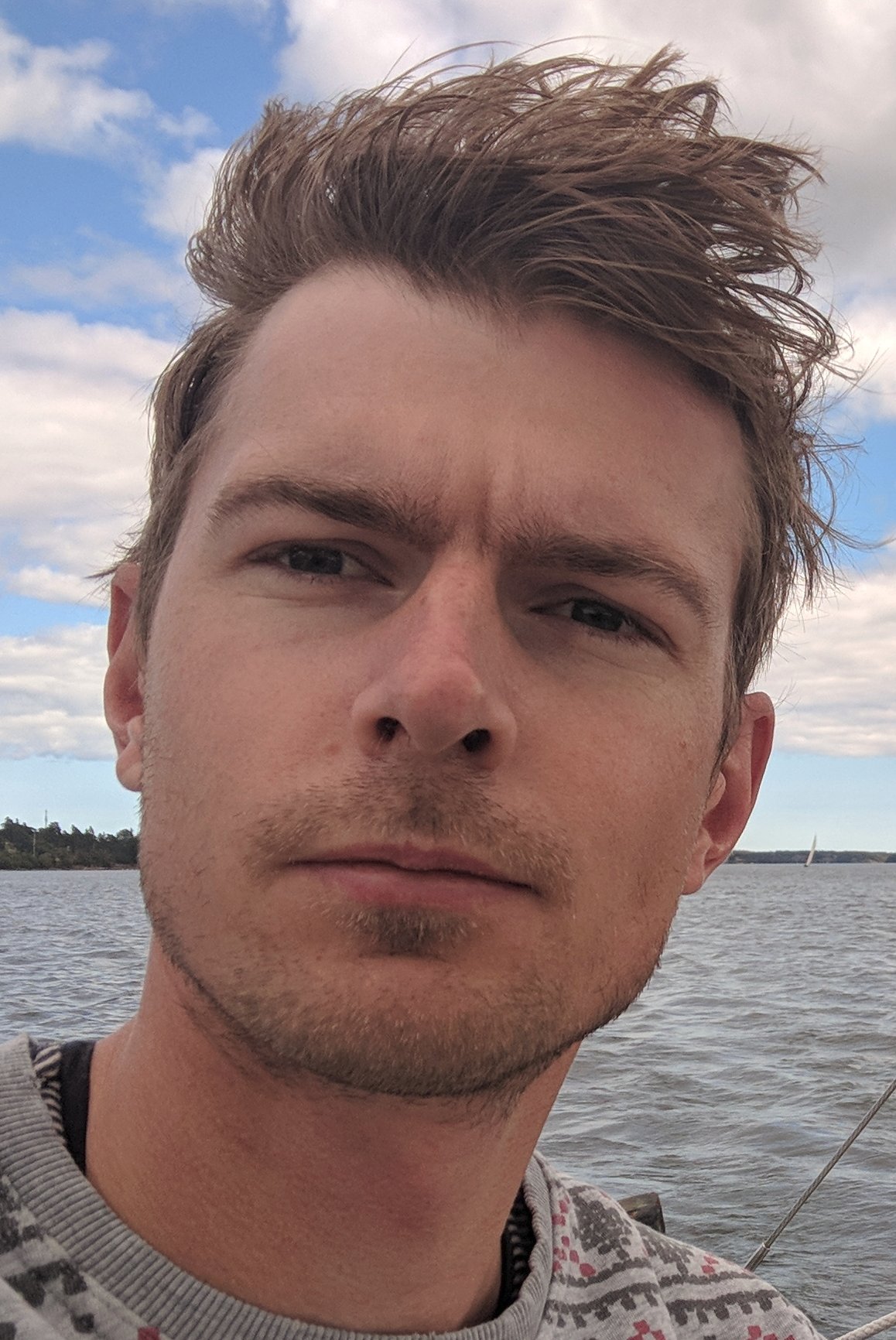}}]{Nils Bore} received the M.Sc. degree in mathematical engineering from the Faculty of Engineering, Lund University, Lund, Sweden, in 2012, and the Ph.D. degree in computer vision and robotics from the Robotics Perception and Learning Lab, Royal Institute of Technology (KTH), Stockholm, Sweden, in 2018.
He is currently a researcher with the Swedish Maritime Robotics (SMaRC) project at KTH. His research interests include robotic sensing and mapping, with a focus on probabilistic reasoning and inference. Most of his recent work has been on applications of specialized neural networks to underwater sonar data. In addition, he is interested in system integration for robust and long-term robotic deployments.
\end{IEEEbiography}


\begin{IEEEbiography}[{\includegraphics[width=1in,height=1.25in,clip,keepaspectratio]{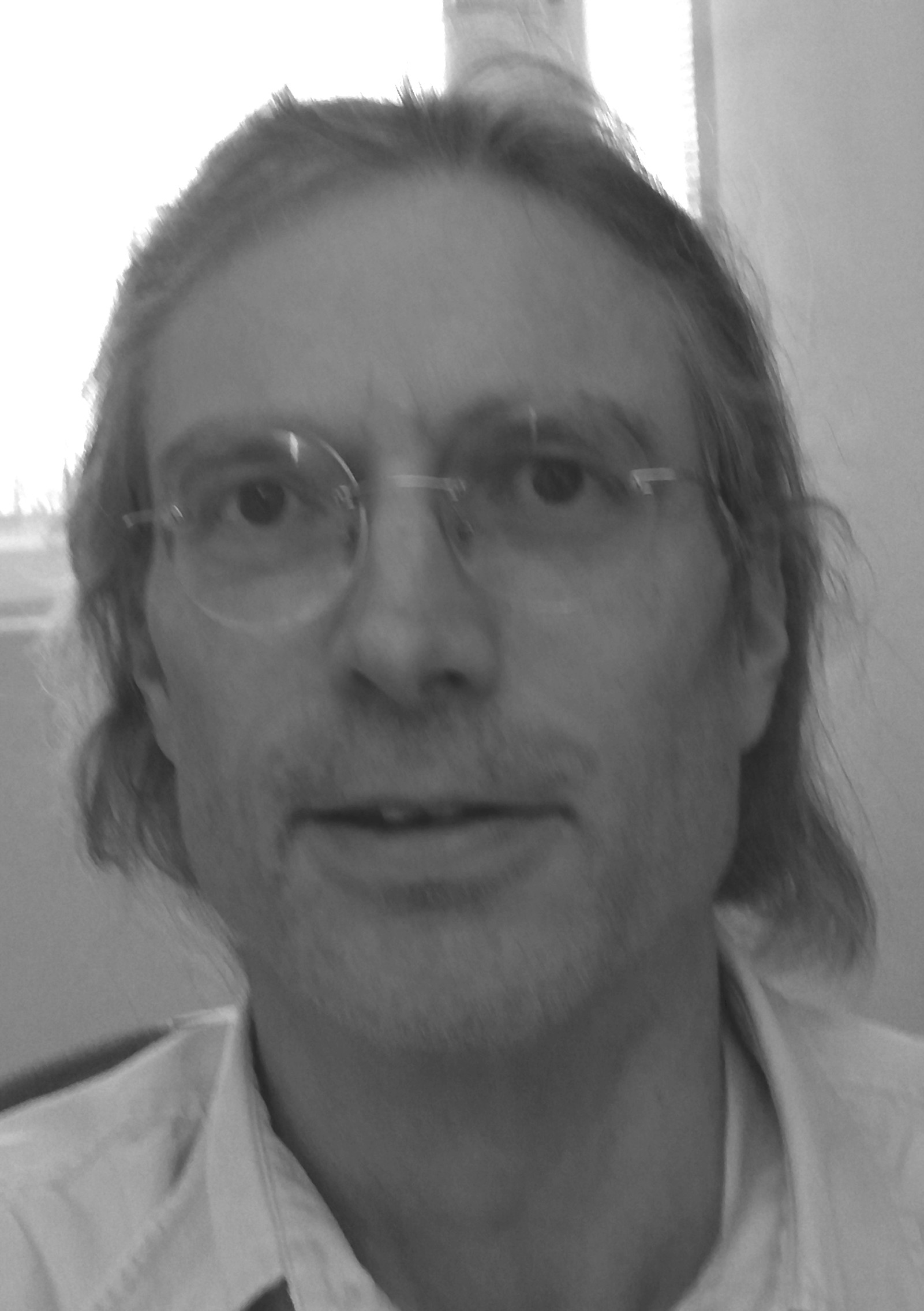}}]{John Folkesson} received the B.A. degree in physics from Queens College, City University of New York,
New York, NY, USA, in 1983, and the M.Sc. degree in computer science, and the Ph.D. degree in robotics
from Royal Institute of Technology (KTH), Stockholm, Sweden, in 2001 and 2006, respectively.
He is currently an Associate Professor of robotics with the Robotics, Perception and Learning Lab, Center for Autonomous Systems, KTH. His research interests include navigation, mapping, perception, and
situation awareness for autonomous  robots.
\end{IEEEbiography}




\end{document}